\documentclass[a4paper,11pt]{article}
\usepackage{pos}
\usepackage{physics,subfigure,cleveref}
\newcommand{\overbar}[1]{\mkern 1.5mu\overline{\mkern-1.5mu#1\mkern-1.5mu}\mkern 1.5mu}
\newcommand{\mn}{{\mu\nu}}
\newcommand{\Tmn}{T_{\mu\nu}}
\newcommand{\TRmn}{T^R_{\mu\nu}}
\newcommand{\rs}{{\rho\sigma}}

\newcommand{\mnrs}{{\mu\nu\rho\sigma}}

\newcommand{\intlm}[1]{\int_{-\frac{\pi}{a}}^{\frac{\pi}{a}}
                       \frac{\diff^3#1}{(2\pi)^3}\,}
\newcommand{\diff}{\mathrm{d}}

\newcommand{\ord}[1]{\mathcal{O}\left({#1}\right)}

\title{Renormalization of the $3D$ $SU(N)$ scalar energy- momentum tensor using the Wilson flow}
\ShortTitle{$3D$ $SU(N)$ scalar EMT renormalization using the Wilson flow}

\author*[a,1]{Joseph~K.~L.~Lee}
\author[a]{Luigi~Del~Debbio}
\author[a,b]{Elizabeth~Dobson}
\author[c,d,e]{Andreas~J\"uttner}
\author[c,d,f]{Ben~Kitching-Morley}
\author[a,g]{Valentin~Nourry}
\author[a]{Antonin~Portelli}%
\author[a]{Henrique~Bergallo Rocha}
\author[d,f]{Kostas~Skenderis}
\affiliation[a]{Higgs Centre for Theoretical Physics, School of Physics and Astronomy, The University of Edinburgh, Edinburgh EH9 3FD, United Kingdom}
\affiliation[b]{Institute of Physics, The University of Graz, Universit\"atsplatz 5, A-8010 Graz, Austria}
\affiliation[c]{School of Physics and Astronomy, University of Southampton, Southampton SO17 1BJ, United Kingdom}
\affiliation[d]{STAG Research Center,
University of Southampton, Highfield, Southampton SO17 1BJ, United Kingdom}
\affiliation[e]{Theoretical Physics Department, CERN, 1211 Geneva 23, Switzerland}
\affiliation[f]{Mathematical Sciences, University of Southampton, Highfield, Southampton SO17 1BJ, United Kingdom}
\affiliation[g]{Universit\'e de Paris, CNRS, Astroparticule et Cosmologie, F-75006 Paris, France}

\note{For the LatCos collaboration.}
\emailAdd{joseph.lee@ed.ac.uk}

\abstract{In the holographic approach to cosmology, cosmological observables are described in terms of correlators of a three-dimensional boundary quantum field theory. As a concrete model, we study the 3D massless SU(N) scalar matrix field theory with a $\phi^4$ interaction. On the lattice, the energy-momentum tensor (EMT) in this theory can mix with the operator $\phi^2$. We utilize the Wilson Flow to renormalize the EMT on the lattice, and present numerical results for the mixing coefficient for $N=2$. Obtaining the renormalized EMT will allow us to make predictions for the CMB power spectra in the regime where the dual QFT is non-perturbative.}

\FullConference{%
 The 38th International Symposium on Lattice Field Theory, LATTICE2021
  26th-30th July, 2021
  Zoom/Gather@Massachusetts Institute of Technology
}

\begin{document}
\maketitle

\section{Introduction}
The energy-momentum tensor (EMT) plays a fundamental role in quantum field theories, by virtue of being the collection of Noether currents related to space-time symmetries. Our main motivation for studying the EMT comes from the application of holography to cosmology~\citep{McFadden:2009fg}. In this holographic approach, cosmological observables, such as the cosmic microwave background (CMB) power spectra, can be described in terms of correlators of the EMT of a dual three-dimensional quantum field theory (QFT) with no gravity. 

This proceeding summarizes the work in~\citep{DelDebbio:2020amx}. In this paper, we are interested in renormalizing the EMT of the simplest version of the holographic dual theories, which is the class of $3d$ massless scalar QFTs with $\phi$ in the adjoin of $SU(N)$ and a $\phi^4$ interaction. 
\section{Scalar Action}
The theory under consideration is a three-dimensional Euclidean scalar $\mathfrak{su}(N)$ valued $\phi^4$ theory,
  \begin{align}
		S\left[ \phi \right] =\frac{a^3N}{g} \sum_{x \in \Lambda^3} \Tr \left[  \sum_\mu \left(\delta_\mu\phi(x)\right)^2 + (m^2-m_c^2) \phi(x)^2 + \phi(x)^4 \right] 
  \end{align}
  with fields $\phi^j_i(x) = \phi^a(x) (T^a)^j_i$, where $\phi^a(x)$ is real, and $T^a$ are the generators of $SU(N)$. Here $\delta_\mu$ is the forward finite difference operator defined by, $\delta_\mu \phi(x) = a^{-1} \left[ \phi(x+a\hat{\mu}) - \phi(x)\right]$, where $\hat{\mu}$ is the unit vector in direction $\mu$, $\Lambda^3$ is a lattice with cubic geometry containing $N_L^3$ points (with periodic boundary conditions), and $a$ the lattice spacing. $g$ is the $\phi^4$ coupling constant with mass dimension one (which does not renormalize), $m^2$ is the bare mass. Since the mass of the theory renormalizes additively, we include the mass counterterm, or \textit{critical mass} $m_c^2(g)$, \textit{i.e.} the value of the bare mass such that the renormalized theory is massless. The critical masses for the ensembles simulated are determined in~\citep{Cossu:2020yeg} at two loops in lattice perturbation theory, as well as nonperturbatively by analyzing the finite-size scaling of the Binder cumulant. The relevant masses are summarized in~\cref{table:critical-mass}.

  \begin{table}[h]
    \centering
    \begin{tabular}{c|c|c|c}
    \hline \hline
     \quad $ag$  \quad &  \quad One loop  \quad &  \quad Two loop  \quad &  \quad Nonperturbative  \quad \\  \hline
     0.1 &  -0.03159 &  -0.03125&   -0.0313408(38) \\  \hline
     0.2 & -0.06318 &  -0.06194 & -0.0622974(98) \\  \hline
     0.3 & -0.09477 & -0.09208  &  -0.092935(16)    \\
     \hline \hline
      \end{tabular}
      \caption{The critical masses $(am_c)^2$ in the infinite volume limit are calculated at NLO in lattice perturbation theory, as well as nonperturbatively in~\citep{Cossu:2020yeg}, which are listed for each 't Hooft coupling $ag$. These are used in the later global fit to obtain $c_3$ in the massless limit.}
      \label{table:critical-mass}
    \end{table}

  \section{Energy-momentum tensor}
  In the continuum theory, the energy-momentum tensor $\Tmn$ is defined as the conserved current of space-time symmetries. For our scalar $SU(N)$ theory, it is given by
	\begin{align}
		\Tmn = \frac{N}{g} \Tr \left\{ 2(\partial_\mu \phi)(\partial_\nu \phi) - \delta_\mn \left[ \sum_\rho (\partial_\rho \phi)^2 + (m^2 - m_c^2) \phi^2 + \phi^4 \right] + \xi \left(\delta_\mn \sum_\rho (\partial_\rho \phi)^2  - (\partial_\mu \phi)(\partial_\nu \phi)\right)\right\}.
\end{align}
Here the term multiplying $\xi$ is the improvement term. In the continuum theory, due to translational invariance, the EMT satisfies Ward-Takahashi identities (WI) of the form 
  \begin{align} \label{eq:continuum-WI}
  	\langle \partial^\mu \Tmn(x) P(y)\rangle = -\bigg \langle \frac{\delta P(y)}{\delta\phi(x)} \partial_\nu \phi(x) \bigg\rangle
  \end{align}
  where $P(y)$ is any composite operator inserted at point $y$. If $P$ is such that the right-hand side of~\cref{eq:continuum-WI} is finite for separated points  $x \neq y$, the left-hand side correlation function, which contains the divergence of the EMT, is finite up to contact terms. For this theory, it can be shown that the insertion of $\Tmn$ does not introduce new UV divergences. The improvement term is identically conserved and trivially satisfies~\cref{eq:continuum-WI}. Therefore $\xi$ will be set to 0 for the remainder of the text.

  On the lattice, the continuous translational symmetry is broken into the discrete subgroup of lattice translations; because of this a na\"ive discretization of the EMT on the lattice,
  \begin{align}
  	\Tmn^0 = \frac{N}{g} \Tr \left\{ 2(\overbar{\delta}_\mu \phi)(\overbar{\delta}_\nu \phi) - \delta_\mn \left[ \sum_\rho (\overbar{\delta}_\rho \phi)^2 + (m^2 - m_c^2) \phi^2 + \phi^4 \right] \right\},
  \end{align}
  which is obtained by replacing the partial derivatives $\partial_\mu \phi(x)$ with the central finite difference $\overbar{\delta}_\mu \phi(x) = \frac{1}{2a}\left[ \phi(x+a\hat{\mu}) - \phi(x-a\hat{\mu}) \right]$ (this is chosen in order to obtain a Hermitian EMT), does not satisfy the WI~\cref{eq:continuum-WI}. Now, the WI on the lattice includes an additional term~\citep{Caracciolo:1988hc},
    \begin{align} \label{eq:lattice-WI}
  	\langle \overbar{\delta}^\mu \Tmn^0 (x) P(y)\rangle = -\bigg \langle \frac{\overbar{\delta}P(y)}{\overbar{\delta}\phi(x)}\overbar{\delta}_\nu \phi(x) \bigg\rangle + \langle X_\nu (x) P(y) \rangle .
  \end{align}
  Here $\frac{\overbar{\delta}P(y)}{\overbar{\delta}\phi(x)}$ is obtained by replacing the fields and derivatives in the continuum functional derivative $\frac{\delta P(y)}{\delta\phi(x)}$ with their lattice counterparts, and $X_\nu$ is an operator proportional to $a^2$, which classically  vanishes in the continuum limit. However, radiative corrections cause the expectation value $\langle X_\nu (x) P(y) \rangle$ to produce a linearly $a^{-1}$ divergent contribution to the WI. Therefore, the na\"ively discretized EMT will not reproduce the continuum WI when the regulator is removed; $\Tmn$ has to be renormalized by adjusting the coefficients of a linear combination of lower-dimensional operators which satisfy the same symmetries.

  In three dimensions, dimensional counting indicates that divergent mixing can only occur with $O_3 = \delta_\mn \Tr \phi^2$. The \textit{renormalized EMT} on the lattice can therefore be defined as an operator mixing,
  \begin{align} \label{eq:renormalised-emt}
  	\TRmn &= \Tmn^0 - c_3 \delta_\mn \frac{N}{a} \Tr \phi^2.
  \end{align}
The coefficient $c_3$ has to be tuned to satisfy the continuum WI up to discretization effects when the regulator is removed.

At leading order (LO) $O(g)$ (i.e. one loop) in lattice perturbation theory, $c_3$ is shown to be
  \begin{align}\label{eq:c3-1loop}
c_3^{\text{1 loop}}&=\left(2-\frac{3}{N^2}\right)\left(\frac{6Z_0-1}{12}\right),  \\
  Z_0 &= a \intlm{k} \frac{1}{\hat{k}^2}  = 0.252731...,
\end{align}   
for lattice momentum $\hat{k} = \frac{2}{a} \sin (ka/2)$. By determining the value of $c_3$ nonperturbatively, we are able to renormalize the EMT on the lattice.

Before discussing the strategy to obtain the value of $c_3$ nonperturbatively, we define an EMT correlator which will be useful in our analysis. Consider the momentum-space two-point correlator,
\begin{align}\label{eq:Tmn0trphi2}
	C_\mn^0(q) = \frac{N}{g} \expval{\Tmn^0 (-q) \Tr \phi^2 (q)}  = \frac{N}{g} a^3 \sum_{x \in \Lambda} e^{-i q \cdot x} \langle \Tmn^0 (x) \Tr \phi^2 (0) \rangle. 
\end{align}
This can be decomposed into three components,
\begin{align} 
  C_\mn^0(q) &= \hat{C}_\mn (q) + c_3 \frac{g}{a} \delta_\mn C_2(q) + \frac{\kappa}{a} \delta_\mn,
\end{align}
where 
\begin{align}
  \hat{C}_\mn (q) &= \frac{N}{g} \left( \expval{\Tmn^R \Tr \phi^2}(q) - \expval{\Tmn^R \Tr \phi^2}(0)\right)\\
  C_2 (q) &= \left(\frac{N}{g}\right)^2  \expval{\Tr \phi^2 \Tr \phi^2}(q)\\
  \kappa^\text{1 loop} &= - \frac{N^2}{2} \left(1-\frac{1}{N^2}\right) \left(\frac{6Z_0 - 1}{12}\right)
\end{align}
The first term $\hat{C}_\mn (q)$ is the renormalized EMT correlator, which can be taken to the continuum limit; the second term $c_3 \frac{g}{a} \delta_\mn C_2(q)$ comes from the divergent operating mixing from~\cref{eq:renormalised-emt}; the third term $\frac{\kappa}{a} \delta_\mn$ comes from a constant (momentum-independent) contact term. 

\section{Wilson Flow}
In order to nonperturbatively renormalize the EMT operator, we need to isolate the contact term from the operator mixing, and we will utilize the method of the Wilson flow~\citep{Luscher:2010iy} to achieve this. For our scalar field $\phi(x)$, define a flowed field $\rho(t,x)$ governed by the flow equations,
\begin{align}
    \partial_t \rho(t,x) = \partial^2 \rho(t,x), \quad \rho(t,x) |_{t=0} = \phi(x),
\end{align}
where $\partial^2 = \sum_\mu \partial_\mu^2$ is the Laplacian, and $t$ is the \textit{flow time}, a new parameter introduced into the theory. Solving by means of Fourier transformation, one finds 
\begin{align}
    \Tilde{\rho}(t, k) = e^{-k^2 t} \Tilde{\phi}(k),
\end{align}
where $\Tilde{\rho}(t,k)$ is the Fourier transform of $\rho(t,x)$; the flow effectively smears the field with radius $\sim\sqrt{t}$.

The Wilson flow suppresses high-momentum modes exponentially, and thereby regulates the divergent contact term present in the EMT correlator $C^0_\mn (q)$. We are therefore able to isolate the divergent mixing $c_3$ from the divergent contact term. 

In our case, we replace the operator $\Tr \phi^2 (x=0)$ in~\cref{eq:Tmn0trphi2} with the operator $\Tr \rho^2(t, x=0)$ at finite flow time $t$, and keep the renormalized EMT operator $\Tmn^R (x)$ at flow time $t=0$ to obtain
\begin{align}
	C_\mn^0(t, q) = \frac{N}{g} \expval{\Tmn^0 (-q) \Tr \rho^2 (t, q)}  = \frac{N}{g} a^3 \sum_{x \in \Lambda} e^{-i q \cdot x} \langle \Tmn^0 (x) \Tr \rho^2 (t, x=0) \rangle. 
\end{align}
By definition, $C^0_\mn(0,q) = C^0_\mn(q)$. Again, this can be decomposed into three components:
\begin{align} \label{eq:small-t-expansion}
  C_\mn^0(t, q) &= \hat{C}_\mn (t,q) + c_3 \frac{g}{a} \delta_\mn C_2(t,q) + K(t) \delta_\mn.
\end{align}
At vanishing flowtime, $K(t=0) = \frac{\kappa}{a}$; however at small finite flowtime,
\begin{align}
  K(t) &= \frac{\omega}{\sqrt{t}} + \ord{\sqrt{t}}\\
  \omega^\text{1 loop} &= -\frac{N^2}{2} \left(1-\frac{1}{N^2}\right)\frac{\sqrt{2}}{24\pi^{3/2}}.
\end{align}
We utilize this small $t$ expansion to remove the contact term contribution in our correlation function in order to obtain the value of $c_3$.

\section{Lattice Setup}
The theory is simulated using the hybrid Monte Carlo algorithm, which was implemented using the Grid library~\citep{Boyle:2016lbp}. For this paper, we focus on the $N=2$ theory. The simulated volumes $N_L^3$, 't Hooft coupling in lattice unit $ag$ (or equivalently the dimensionless lattice spacing), and bare masses $(am)^2$ are listed in~\cref{tabel:ensemble}. For each of the three 't Hooft couplings, two bare masses in the vicinity of the critical mass have been simulated (see~\cref{table:critical-mass}).
\begin{table}[h]
  \begin{minipage}{.5\linewidth}
    \centering
      \begin{tabular}{c|c}
      \hline \hline
         \quad $ag$  \quad &  \quad $(am)^2$ \quad \\ \hline
           0.1 & -0.0305, -0.031 \\
           0.2 & -0.061, -0.062 \\ 
           0.3 & -0.092, -0.091 \\ 
           \hline \hline
      \end{tabular}
  \end{minipage}%
  \begin{minipage}{.5\linewidth}
    \centering
      \begin{tabular}{c|c|c}
      \hline \hline
           \quad $N_L^3$  \qquad & \quad Trajectories \quad & \quad Sample frequency \\ \hline
          $64^3$ & 1,500,000 & 50\\ 
          $128^3$& 500,000 & 50\\
          $256^3$ & 200,000 & 100\\
          \hline \hline
      \end{tabular}
  \end{minipage} 
  \caption{For each 't Hooft coupling $ag$, two bare masses are simulated in three volumes}
  \label{tabel:ensemble}
\end{table}

\section{EMT renormalization condition}
The renormalization scheme is defined by imposing the Ward identity
\begin{align} \label{eq:lattice-WI-renorm-condition}
\overbar{q}_\mu \hat{C}_\mn (t, q) = 0
\end{align}
on all lattice ensembles. Here $\overbar{q}=\frac{1}{a} \sin \left(aq\right)$ is the lattice momentum. This condition is imposed on specific values of momentum $aq^*$. This gives a value of $c_3$ for each choice of momentum, mass, volume and 't Hooft coupling. We then extrapolate the value $c_3$ towards the massless and infinite volume limit to obtain $\overbar{c_3}$. This defines a massless renormalization scheme, which is independent of the volume.

The renormalization condition~\cref{eq:lattice-WI-renorm-condition} implies that $\hat{C}_\mn (t,q)$ is purely transverse, \textit{i.e.}, 
\begin{align} \label{eq:Cmntq-transverse}
  \hat{C}_\mn (t,q) \propto \overbar{\pi}_\mn,  
\end{align}
where $\overbar{\pi}_\mn = \delta_\mn - \frac{\overbar{q}_\mu\overbar{q}_\nu}{\overbar{q}^2}$ is the transverse projector. In other words, for a fixed longitudinal momentum $q_l^* = (0,0,q^*)$, $\hat{C}_{22} (t, q^*_l) = 0$.

By rearranging~\cref{eq:small-t-expansion} for the longitudinal component, we obtain the expression
\begin{align} \label{eq:c3-fit-form}
  \frac{a}{g}\frac{C^0_{22} (t, q_l^*)}{C_2 (t, q_l^*)} = c_3 + \frac{\Omega}{g \sqrt{t}}\ , \ \text{where} \quad \Omega = \frac{\sqrt{2} a q_l^*}{3\pi^{3/2}}.
\end{align}

We perform a fit using a linear function of the \textit{inverse physical flowtime} $\frac{1}{g \sqrt{t}}$, leaving $\Omega$ and $c_3$ as fit parameters. From the fit, we can extrapolate $c_3$ from the y-intercept.

\section{Numerical Results}
\begin{figure}[ht]
  \centering
  \subfigure[\ $ag=0.1, N_L=64$]{
  \includegraphics[scale=0.75]{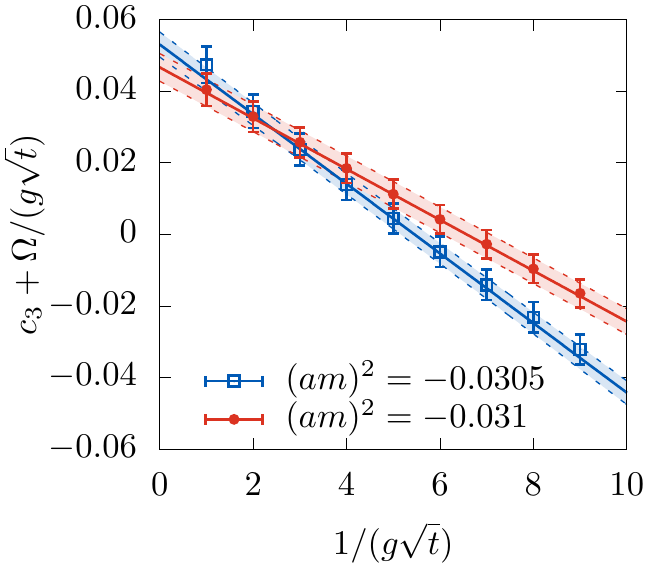}}
  \subfigure[\ $ag=0.1, N_L=128$]{
  \includegraphics[scale=0.75]{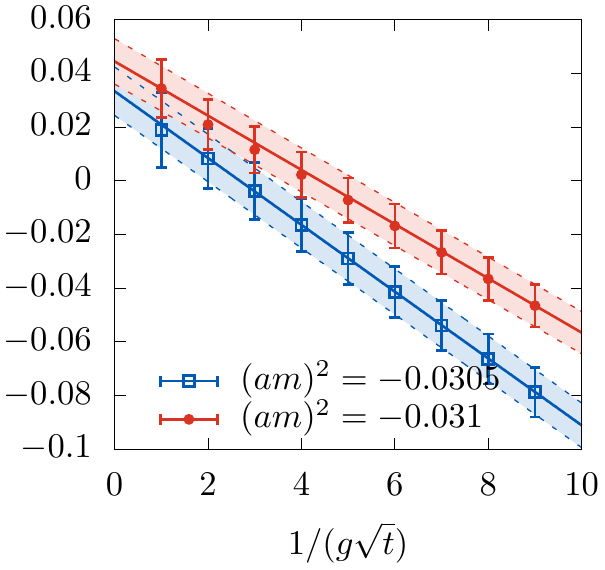}}
  \subfigure[\ $ag=0.1, N_L=256$]{
  \includegraphics[scale=0.75]{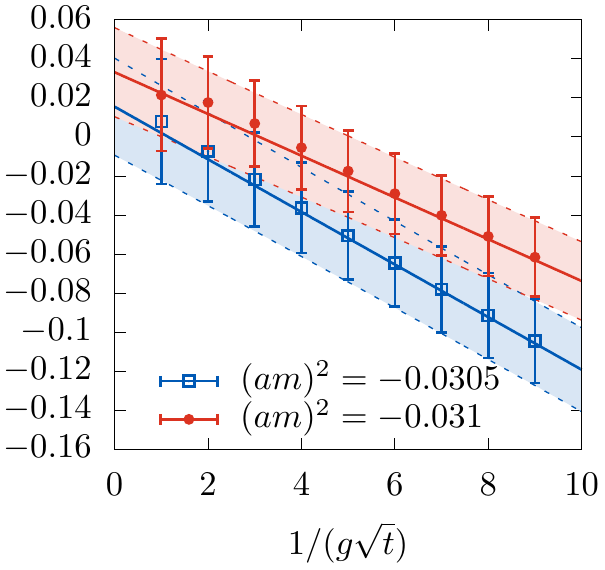}}
  \caption{Plots showing $c_3$ against the inverse physical flow time $\frac{1}{g\sqrt{t}}$ using~\cref{eq:c3-fit-form} for momentum $a\vert q^*_l \vert = 0.098$ with 't Hooft coupling $ag=0.1$ for three volumes. The red and blue data points are for the lighter and heavier mass simulations respectively, and the corresponding error bands in the fit are from statistical uncertainty. The value of $c_3$ is the $y$ intercept on the fit.\label{fig:c3fit-plot}}
  \end{figure}

Picking the fit ranges for the physical flow time $g\sqrt{t}$ requires special attention. They must first be sufficiently small to justify the small flow time expansion of~\cref{eq:small-t-expansion}. This also ensures the smearing radius is sufficiently smaller than the length of the lattice ($gL=gaN_L$) such that there will be small finite volume contributions from the boundaries. The physical flow time must also be larger than the lattice spacing ($ag$) such that actual smearing occurs across lattice points. We therefore impose the range to be between $ag < g \sqrt{t} < 1$. We performed the analysis for four values of momenta $a\vert q^*_l\vert = 0.049$, $0.098$, $0.147$, $0.196$. Some examples of the fit are shown in~\cref{fig:c3fit-plot}, the full fit results can be found in~\citep{DelDebbio:2020amx}.

  In order to include the mass, volume and lattice-spacing dependence of the value of $c_3$, we perform global fits using
  \begin{align} \label{eq:global-fit-model}
      c_3(\overbar{m^2_R},gL, ag) = \overbar{c_3} + p_0 \overbar{m^2_R} + \frac{p_1}{gL} + p_2 (ag),
  \end{align}
  where $\overbar{m^2_R}=(m^2-m_c^2)/g^2$ is the dimensionless renormalized mass (The values of $m_c^2$ are summarized in~\cref{table:critical-mass}), $gL$ is the dimensionless length of the lattice, and $ag$ the dimensionless lattice spacing. As we have chosen our simulation to have large volume, small lattice spacing, and close to the critical mass, we believe that the linear corrections are appropriate. In particular, since the divergent mixing is a UV effect, we expect there to be small volume dependence coming from the IR. 
  
\begin{figure}[ht]
    \centering
    \subfigure[\ Model 2 (against $\overbar{m_R^2}$)]{
    \includegraphics[scale=0.85]{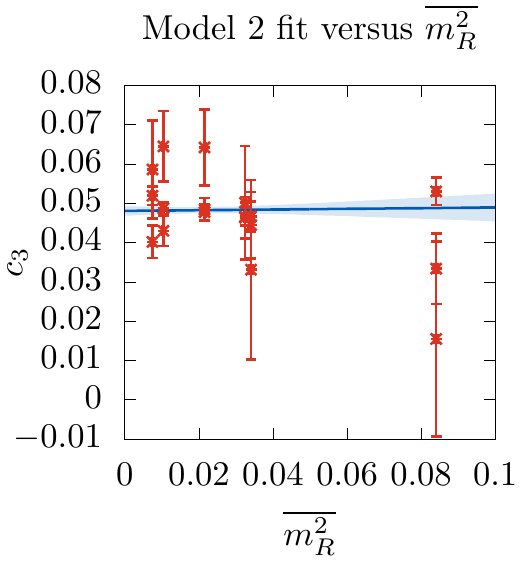}}
    \subfigure[\ Model 3 (against $gL$)]{
    \includegraphics[scale=0.85]{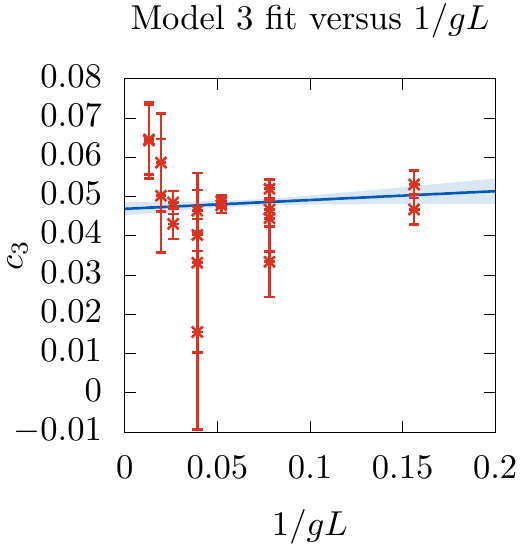}}
    \subfigure[\ Model 4 (against $ag$)]{
    \includegraphics[scale=0.85]{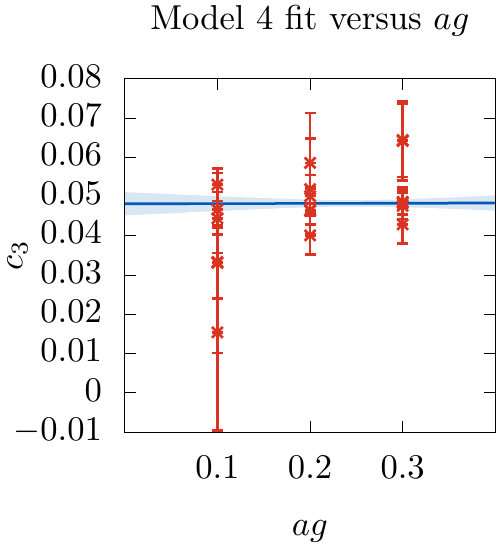}}
    \caption{$c_3$ global fits using model 2, 3, 4 for $a \vert q^*_l \vert = 0.098$. Each plot is plotted against the respective free fitting parameter for each model, \textit{i.e.} $\overbar{m_R^2}$, $gL$, and $ag$; the value for $\overbar{c_3}$ is the $y$-intercept of the fit line.\label{fig:global-fits}}
  \end{figure}

  For the global fits, the three parameters $p_0$, $p_1$, $p_2$ are switched on individually, resulting in $2 \times 2 \times 2 = 8$ fit models for each of the four momenta, which gives a total of 32 fit results for the value of $\overbar{c_3}$. \cref{fig:global-fits} shows examples of the global fits for momentum $a \vert q^*_l \vert = 0.098$.

\begin{figure}[ht]
  \centering
  \subfigure[The values of $\overbar{c_3}$ for models with no fit parameters which are $0.5 \sigma$ compatible with 0.]{
  \includegraphics[scale=0.6]{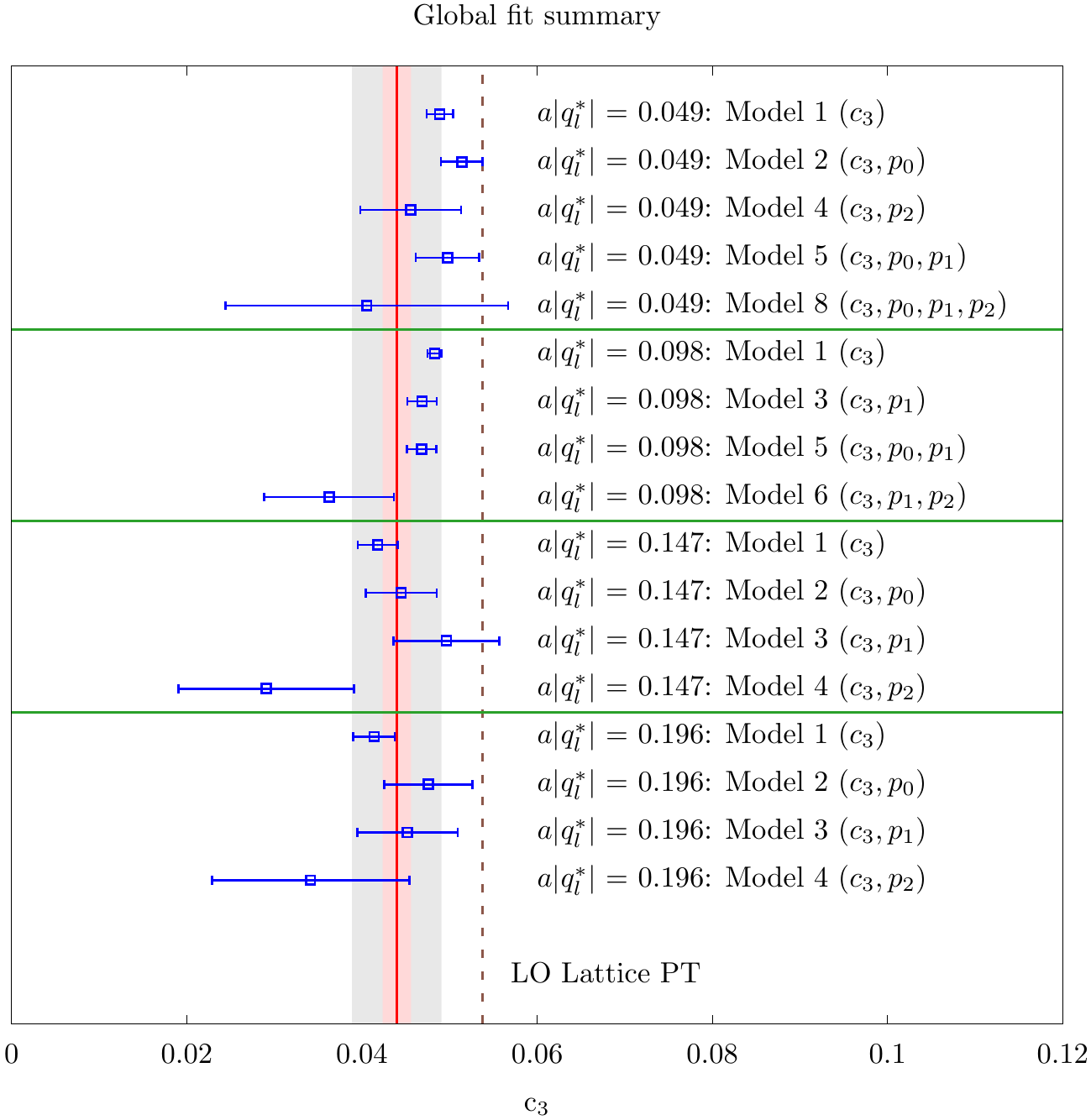}}
  \subfigure[Histogram of the distribution of $\overbar{c_3}$.]{
  \includegraphics[scale=0.7]{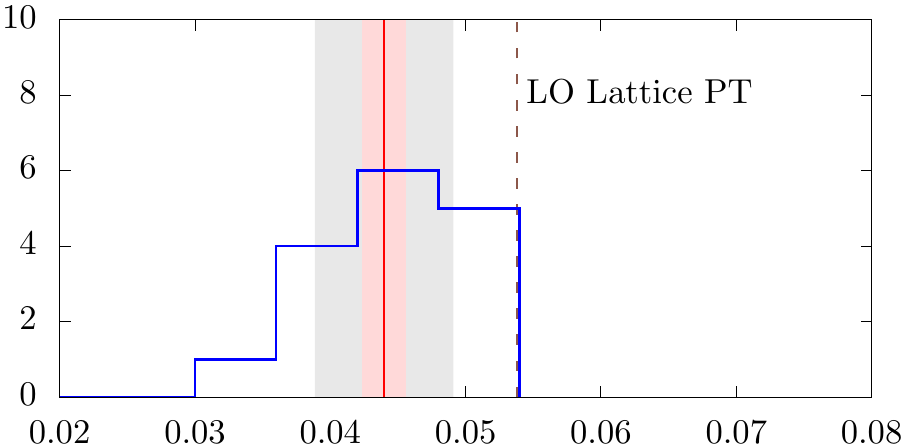}}
  \caption{The red line shows the final central result, the red and gray bands represent the statistical and systematic errors respectively. The brown dashed line shows the one-loop perturbation theory value from~\cref{eq:c3-1loop}. The fact that the nonperturbative result is close to the one-loop perturbative value is expected due to the super-renormalizability of the theory.}
  \label{fig:global-fit-summary}
\end{figure}

  In order to estimate the final statistical and systematic errors, we adopt the following procedure inspired by~\citep{Durr:2008zz}. We construct the distribution of values for $\overbar{c_3}$ from global fits which does not include any parameter with a fit value $0.5 \sigma$ compatible with 0. From the 17 results within the distribution, the central value of $\overbar{c_3}$ is defined to be the mean of the distribution, the statistical error to be the statistical error of the mean as measured with the bootstrap samples, and the systematic error to be the symmetrized central 68.3\% confidence interval of the distribution.
  A summary of the values of $\overbar{c_3}$ and a histogram of the distribution are shown in~\cref{fig:global-fit-summary}, along with the one-loop value $c_3^{\text{1 loop}}$ from~\cref{eq:c3-1loop}. This procedure yields the final result $\overbar{c_3} = 0.0440(16)_\text{stat}(51)_\text{sys}$.
  
It is worth noting that the finiteness of this value in the infinite volume limit is a nonpeturbative feature of the theory. In perturbation theory, all terms of $\ord{g^2}$ are IR divergent and depend on the IR regulator; but as shown in~\citep{Cossu:2020yeg} the theory is in fact nonperturbatively IR finite, where the dimensionful coupling effectively acts as the IR regulator in the infinite volume limit. Comparing the nonperturbative result for $\overbar{c_3}$ with the one-loop perturbative value, the nonperturbative value is approximately $20\%$ smaller than the one-loop result. This is qualitatively expected, as the higher order terms in perturbation theory (with the IR regulator replaced by the coupling) changes sign at every order, and the two-loop result is a correction of the opposite sign to the one-loop value.

\section{Conclusion and Outlook}
We have presented a procedure to nonperturbatively renormalize the EMT on the lattice for a three-dimensional scalar QFT with a $\phi^4$ interaction and field $\phi$ in the adjoint of $SU(N)$. We have also presented numerical results of the EMT operator mixing for the theory with $N=2$. The method utilizes the Wilson flow to define a probe at positive flow time, which can eliminate the divergent contact term present in the EMT correlator. This allows us to determine the mixing coefficient with the lower-dimensional operator $\delta_\mn\Tr \phi^2$. This ensures that the Ward identity can be restored in the continuum limit, up to cutoff effects.

The context of our investigation is to predict the CMB power spectrum for holographic cosmological models, and to test them against observational data. The next step of the investigation is to determine the renormalized EMT two-point function, $C_\mnrs (q) = \langle T_\mn(q) T_\rs(-q) \rangle$, for this class of scalar theories. This two-point function can be used to compute the primordial CMB power spectra in the holographic cosmology framework. On the lattice, this correlator contains a large contact term of order $\ord {a^{-3}}$. This large contact term presents significant statistical noise to the signal of the renormalized two-point function. We are currently exploring using a position-space-based filtering technique to eliminate the presence of such a contact term, which will allow us to make a fully nonperturbative prediction for the CMB power spectra with the $SU(N)$ scalar theory as the dual theory.

We are also working towards simulating and performing the renormalization of the EMT for three-dimensional QFTs with adjoint $SU(N)$ scalars coupled to gauge fields. This is the class of theories preferred by the fit of the perturbative predictions to Planck data~\citep{Afshordi:2016dvb}. In these theories, the lattice EMT contains more counterterms which need to be determined. Much work has been performed in studying the EMT on the lattice for gauge theories~\citep{Caracciolo:1991cp}. The implementation of the Wilson flow for renormalizing the EMT has also been studied for gauge theories, e.g.~\citep{Suzuki:2013gza}. We are exploring related methods to perform renormalization of the EMT for theories with scalar fields coupled to gauge fields. This will take us closer to fully testing the viability of holographic cosmological models as a description of the very early Universe.

\section*{Acknowledgments}
  The authors would like to warmly thank Pavlos Vranas for his valuable support during the early stages of this project. We thank Masanori Hanada for collaboration at initial stages of this project. Simulations produced for this work were performed using the Grid~\citep{Boyle:2016lbp} and Hadrons libraries, which are free software under GPLv2. The data analysis was based on the LatAnalyze library, which is free software under GPLv3. This work was performed using the Cambridge Service for Data Driven Discovery (CSD3), part of which is operated by the University of Cambridge Research Computing on behalf of the STFC DiRAC HPC Facility. The DiRAC component of CSD3 was funded by BEIS capital funding via STFC capital grants ST/P002307/1 and ST/R002452/1 and STFC operations grant ST/R00689X/1. DiRAC is part of the National e-Infrastructure. A.J. and K.S. acknowledge funding from STFC consolidated grants ST/P000711/1 and ST/T000775/1. A.P. is supported in part by UK STFC grant ST/P000630/1. A.P., J.K.L.L., V.N., and H.B.R are funded in part by the European Research Council (ERC) under the European Union’s Horizon 2020 research and innovation programme under grant agreement No 757646 and A.P. additionally grant agreement No 813942. J.K.L.L. is also partly funded by the Croucher foundation through the Croucher Scholarships for Doctoral Study. B.K.M. was supported by the EPSRC Centre for Doctoral Training in Next Generation Computational Modelling Grant No. EP/L015382/1. V.N. is partially funded by the research internship funds of the Universit\'e Paris-Saclay. L.D.D. is supported by an STFC Consolidated Grant, ST/P0000630/1, and a Royal Society Wolfson Research Merit Award, WM140078. 

\bibliographystyle{JHEP}
\bibliography{cosmhol-emt-pos}

\end{document}